\documentclass[reprint,amssymb,aps,prb]{revtex4-2}

\usepackage{CJK}
\usepackage{amsfonts, amssymb, amsmath, amsthm, mathtools}
\usepackage{physics}
\usepackage{graphicx}
\usepackage{xcolor}
\usepackage[caption=false]{subfig}
\newcommand{\phantomsubcaption}[1]{
    {
        \captionsetup[subfloat]{farskip=0pt,captionskip=0pt}
        \captionsetup[subfigure]{labelformat=empty}
        \subfloat{#1}
    }%
}
\usepackage{hyperref}
\hypersetup{
    colorlinks=true,
    linkcolor=blue,
    filecolor=magenta,      
    urlcolor=cyan,
    }
\usepackage{cleveref}
\usepackage[inline]{enumitem}
\usepackage{verbatim}
\RequirePackage[normalem]{ulem} 
\RequirePackage{color}\definecolor{RED}{rgb}{1,0,0}\definecolor{BLUE}{rgb}{0,0,1} 
\providecommand{\DIFdeltex}[1]{}                      
\providecommand{\DIFaddbegin}{} 
\providecommand{\DIFaddend}{} 
\providecommand{\DIFdelbegin}{} 
\providecommand{\DIFdelend}{} 
\providecommand{\DIFaddbeginFL}{} 
\providecommand{\DIFaddendFL}{} 
\providecommand{\DIFdelbeginFL}{} 
\providecommand{\DIFdelendFL}{} 
\newcommand{\DIFscaledelfig}{0.5}
\RequirePackage{settobox} 
\RequirePackage{letltxmacro} 
\newsavebox{\DIFdelgraphicsbox} 
\newlength{\DIFdelgraphicswidth} 
\newlength{\DIFdelgraphicsheight} 
\LetLtxMacro{\DIFOincludegraphics}{\includegraphics} 
\newcommand{\DIFaddincludegraphics}[2][]{{\color{blue}\fbox{\DIFOincludegraphics[#1]{#2}}}} 
\newcommand{\DIFdelincludegraphics}[2][]{
\sbox{\DIFdelgraphicsbox}{\DIFOincludegraphics[#1]{#2}}
\settoboxwidth{\DIFdelgraphicswidth}{\DIFdelgraphicsbox} 
\settoboxtotalheight{\DIFdelgraphicsheight}{\DIFdelgraphicsbox} 
\scalebox{\DIFscaledelfig}{
\parbox[b]{\DIFdelgraphicswidth}{\usebox{\DIFdelgraphicsbox}\\[-\baselineskip] \rule{\DIFdelgraphicswidth}{0em}}\llap{\resizebox{\DIFdelgraphicswidth}{\DIFdelgraphicsheight}{
\setlength{\unitlength}{\DIFdelgraphicswidth}
\begin{picture}(1,1)
\thicklines\linethickness{2pt} 
{\color[rgb]{1,0,0}\put(0,0){\framebox(1,1){}}}
{\color[rgb]{1,0,0}\put(0,0){\line( 1,1){1}}}
{\color[rgb]{1,0,0}\put(0,1){\line(1,-1){1}}}
\end{picture}
}\hspace*{3pt}}} 
} 
\LetLtxMacro{\DIFOaddbegin}{\DIFaddbegin} 
\LetLtxMacro{\DIFOaddend}{\DIFaddend} 
\LetLtxMacro{\DIFOdelbegin}{\DIFdelbegin} 
\LetLtxMacro{\DIFOdelend}{\DIFdelend} 
\DeclareRobustCommand{\DIFaddbegin}{\DIFOaddbegin \let\includegraphics\DIFaddincludegraphics} 
\DeclareRobustCommand{\DIFaddend}{\DIFOaddend \let\includegraphics\DIFOincludegraphics} 
\DeclareRobustCommand{\DIFdelbegin}{\DIFOdelbegin \let\includegraphics\DIFdelincludegraphics} 
\DeclareRobustCommand{\DIFdelend}{\DIFOaddend \let\includegraphics\DIFOincludegraphics} 
\LetLtxMacro{\DIFOaddbeginFL}{\DIFaddbeginFL} 
\LetLtxMacro{\DIFOaddendFL}{\DIFaddendFL} 
\LetLtxMacro{\DIFOdelbeginFL}{\DIFdelbeginFL} 
\LetLtxMacro{\DIFOdelendFL}{\DIFdelendFL} 
\DeclareRobustCommand{\DIFaddbeginFL}{\DIFOaddbeginFL \let\includegraphics\DIFaddincludegraphics} 
\DeclareRobustCommand{\DIFaddendFL}{\DIFOaddendFL \let\includegraphics\DIFOincludegraphics} 
\DeclareRobustCommand{\DIFdelbeginFL}{\DIFOdelbeginFL \let\includegraphics\DIFdelincludegraphics} 
\DeclareRobustCommand{\DIFdelendFL}{\DIFOaddendFL \let\includegraphics\DIFOincludegraphics} 

\begin{document}
\begin{CJK*}{UTF8}{}
\title{
Strange metal phase of disordered magic-angle twisted bilayer graphene at low temperatures: from flatbands to weakly coupled Sachdev-Ye-Kitaev bundles
}
\author{Chenan Wei (\CJKfamily{gbsn}魏晨岸)}
\author{Tigran A. Sedrakyan}
\affiliation{Department of Physics, University of Massachusetts, Amherst, Massachusetts 01003, USA}
\date{\today}
\begin{abstract}
We use stochastic expansion and exact diagonalization to study the magic-angle twisted bilayer graphene (TBG) on a disordered substrate. We show that the substrate-induced strong Coulomb disorder in TBG with the chemical potential at the level of the flatbands drives the system to a network of weakly coupled Sachdev-Ye-Kitaev (SYK) bundles, stabilizing an emergent quantum chaotic strange metal (SM) phase of TBG that exhibits the absence of quasiparticles. The Gaussian orthogonal ensemble dominates TBG's long-time chaotic dynamics at strong disorder, whereas fast quantum scrambling appears in the short-time dynamics. In weak disorder, gapped phases of TBG exhibit exponentially decaying specific heat capacity and exponential decay in out-of-time-ordered correlators (OTOC). This is the system behavior in correlated insulator and superconducting phases, in agreement with the corresponding Larkin-Ovchinnikov result for correlators.
The result suggests a low-temperature transition from the superconducting and correlated insulating phases into the strange metal upon increasing the disorder strength. We propose a finite-temperature phase diagram for Coulomb-disordered TBG and discuss the experimental consequences of the emergent SM phase.

\end{abstract}
\maketitle
\end{CJK*}


\section{Introduction}

The discovery of superconductivity and correlated insulating states in twisted bilayer graphene (TBG) samples\cite{cao2018unconventional,cao2018correlated}, where tunneling spectra show the superconductivity, an unconventional nodal superconductor phase\cite{oh2021evidence}, has led to increased interest in studying the impact of flatbands in TBG resulting from ``magic" twist angles\cite{bistritzer2011moire,tarnopolsky2019origin}. The presence of non-dispersive bands enhances the effect of electron correlations making TBG an ideal venue to study the emergent phenomena stabilizing unconventional phases of strongly correlated electrons \cite{kerelsky2019maximized,nuckolls2020strongly,wong2020cascade,lisi2021observation,jaoui2022quantum}.
Recent extensive studies of TBG's correlated many-body properties have uncovered various interesting phenomena, including flatband ferromagnetism\cite{pons2020flat}, $SU(2)$ collective excitations\cite{pan2021dynamic}, Kohn-Luttinger instability\cite{gonzalez2019kohn}, near-Planckian dissipation\cite{polshyn2019large}.
The effect of the angular disorder is especially interesting, shown to smear the flatband and enhance the transmission\cite{wilson2020disorder,padhi2020transport,joy2020transparent,thomson2021recovery,sainz2021high}.

\begin{figure}
    \centering
    \begin{minipage}{0.9\linewidth}
    \includegraphics[width=\textwidth]{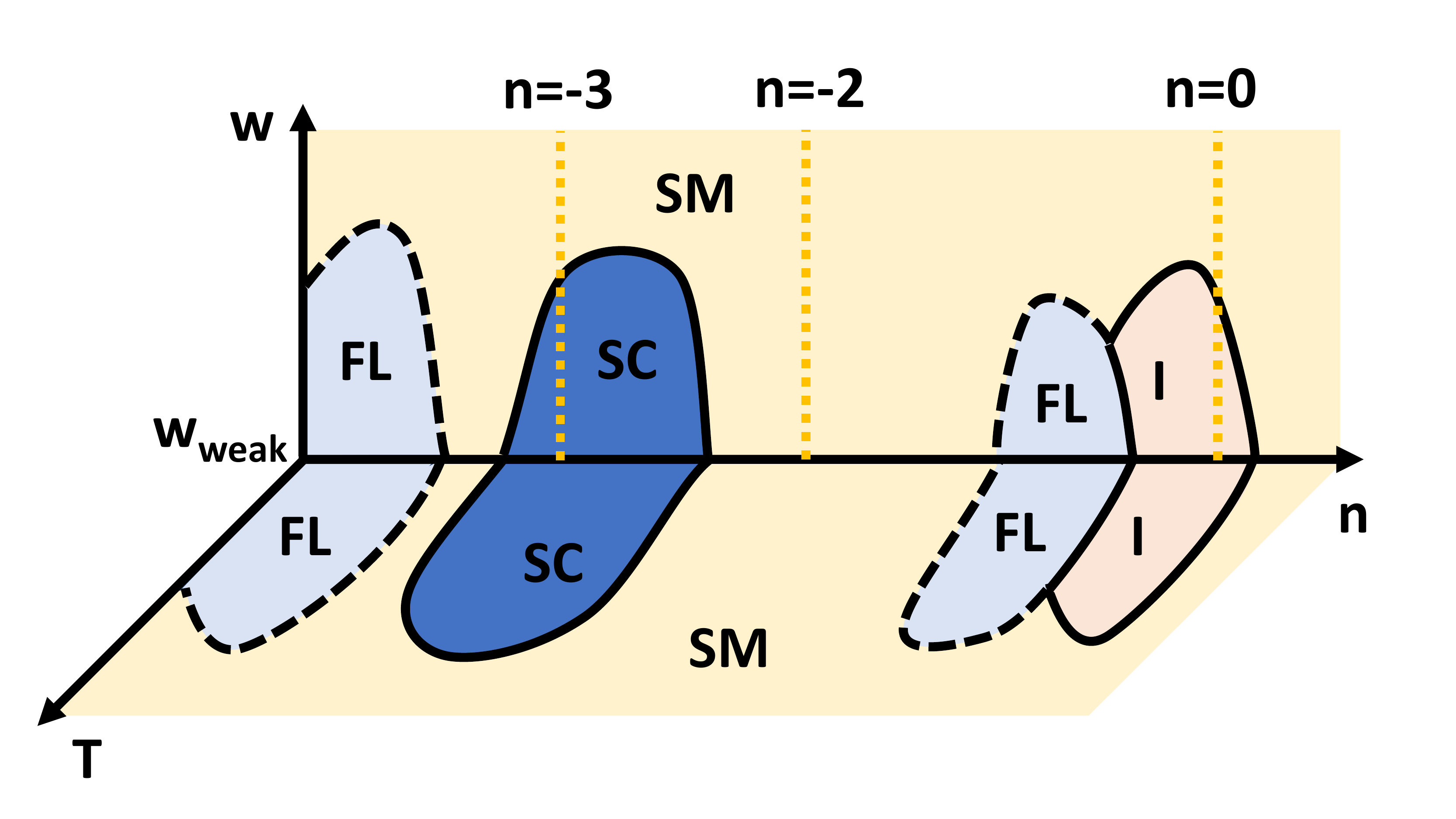}
    \end{minipage}
    \caption{\label{phase}
The schematic phase diagram of DTBG.
``SC" represents the superconducting phase, ``SM" is the quantum chaotic strange metal phase, ``I" is the correlated insulating phase, and ``FL" is the Fermi liquid phase.
The orange dashed lines, labeled by $n$, represent the number of electrons per Mori\'e unit cell with respect to the charge neutrality point used in numerical simulations.
At $n = -3$ and $n = 0$, chaotic SM is observed at strong disorder, while gapped phases emerge at weak disorder for these electron densities. When $n = -2$, the SM phase persists even at weak disorder.
In the figure, the origin of the $w$ axis is a finite weak disorder scale $w_{\rm weak}\lesssim 10^{-3}t$ where $t$ is intralayer hopping integral.
In experiments, even clean samples exhibit a small degree of disorder, which still allows for the persistence of the SM phase as observed experimentally\cite{cao2020strange,jaoui2022quantum}.
We predict that the system undergoes an SM behavior even at low temperatures upon increasing the disorder strength.
}
\end{figure}

However, the effects of residual impurities and smooth Coulomb disorder in such correlated flatband systems\cite{daniel2018artificial} and the interplay with strong electron-electron interactions are poorly understood. The emergence of quantum chaotic Sachdev-Ye-Kitaev (SYK)-like phase, a maximum chaotic non-Fermi liquid (non-FL) with holographic dual\cite{sachdev1993gapless,maldacena2016remarks,garcia2016spectral,davison2017thermoelectric,garcia2017analytical,sedrakyan2020supersymmetry,garcia2021sparse,jia2022replica,garcia2023sixfold,bandyopadhyay2023universal}, is one of the outcomes if point-like impurities are present in the flatband systems such as in the quantum matter with kagome lattice structure\cite{wei2021optical}. The geometry of such line-graph lattice protects the flatbands\cite{chiu2020fragile}. However, the origin of 
flatbands in magic-angle TBG is different: the chiral symmetry of the Moir\'e superlattice enables the emergence of the dispersionless localized flatbands\cite{tarnopolsky2019origin}. 
The collapse of Van Hove singularities provides another intuitive understanding of the flatbands \cite{yuan2019magic,wu2021chern,tilak2021flat}.
The Van Hove singularities of two overlapping Dirac cones appear at energies $\pm \hbar v_0 k_{\theta}$, where $v_0$ is the Fermi velocity of the single-layer graphene, $k_{\theta} = 2 k_D \sin (\theta/2)$ and $k_D$ is the magnitude of the wave vector at the corner of the Brillouin-zone for a single layer.  The Van Hove singularities will be perturbed and extended by the interlayer hopping $t'$. When $\hbar v_0 k_{\theta} \sim 2t'$, namely $\theta \sim \frac{2t'}{\hbar v_o k_{D}}$, the nearly flatbands will appear, which can be regarded flat within an energy interval of order of 10 meV\cite{cao2018unconventional} (throughout this article we will refer to this energy interval implying flatness of the band as ``small''). The flat band states are localized within the Moir\'e cell. The direct result of this localization is that the overlaps between the flat band states become sparse. Interestingly, compared to the disordered kagome lattice, this effect leads to different thermodynamics of such disordered TBG (DTBG).

The dynamics, unlike thermodynamics, is dominated by the local structure of states' overlaps and is described by the out-of-time-order correlator ($OTOC$).
For quantum scrambling dynamics, $OTOC \equiv \langle X(x, \tau) Y(0, 0) X(x, \tau) Y(0, 0) \rangle_{\beta}$ is the exponential growing part of the commutator $- \langle [ X(x, \tau), Y(0, 0) ]^2 \rangle_{\beta}$, where $X(x, \tau)$ and $Y(0, 0)$ are operators in Heisenberg picture and $\langle...\rangle_\beta$ is the regularized thermal average at inverse temperature $\beta$, describing the rate of local information spreading over the entire system in the semi-classical limit. Namely, such commutator of a pair of canonical variables is proportional to the $\left( \frac{\partial X(x, \tau)}{\partial Y(0, 0)} \right)^2$ in the semi-classical limit. Inspired by the gravity duality, 
the Lyapunov exponent $\lambda$ of the correlator, \textit{i.e.}, $OTOC \sim e^{\lambda (\tau - x/v_B)}$(where $v_B$ is the butterfly velocity\cite{han2019quantum}) is proposed to have a maximal bound $\lambda \le \lambda_{max}\equiv 2\pi T$\cite{maldacena2016bound,gu2019relation}. Even though the upper bound coefficient can be different
in the absence of dilatons and when the dual AdS spacetime is close to the flat spacetime\cite{gwak2022violation},
the linear dependence on the temperature is still strong evidence for quantum scrambling. Importantly, the Lyapunov exponent $\lambda$ scales as $T^2$ for FL at low temperatures. $\lambda$ can generally scale as $T^{\gamma}$, where $1 \le \gamma \le 2$ results from the quantum criticality and many-body effects. The lower bound $\gamma = 1$ evidences the lacking of long-lived quasiparticles\cite{hartnoll2021planckian,tikhanovskaya2022maximal,sedrakyan2009fermionic}. In gapped phases, namely superconducting and correlated insulating phases, the semi-classical limit of $OTOC$ is shown to decay exponentially in $T$\cite{larkin1969quasiclassical}.

The present work studies the behavior of $OTOC$ in Coulomb disordered magic-angle twisted TBG to pinpoint the region where superconductivity is stabilized\cite{lantagne2021superconducting,klebanov2020spontaneous,wang2020sachdev,classen2021superconductivity,inkof2021quantum} and aims at understanding the nature of the phase outside the superconducting region. Our analysis suggests the emergence of SYK physics in TBG and establishes a connection with superconductivity. Our central finding is that in a broad region of temperatures and disorder strengths, the TBG system exhibits the emergent weakly and randomly coupled SYK granules, stabilizing a strange metal (SM) phase. 
Our approach proposes a novel scenario of superconductivity in TBG based on the flat band and disorder-induced SYK bundles.
Moreover, our theory proposes a finite-temperature phase diagram depicted in \cref{phase}. This is especially important because superconductivity is experimentally observed not in most available magic-angle TBG samples but 
only in some of them.
 To this end, the suggested phase diagram can be a systematic tool for characterizing the behavior of the TBG samples based on the disorder strength and corresponding observable characteristics.

One type of disorder in TBG is associated with the slight variation of twist angle in different regions of two-dimensional samples. This type of disorder leads to certain interesting renormalization effects discussed in detail in Refs.~ \cite{wilson2020disorder,padhi2020transport,joy2020transparent,thomson2021recovery}. On the other hand, the angular disorder can be kept under experimental control, so we do not consider its consequences in the present work.

\begin{figure}
    \centering
    \begin{minipage}{0.49\linewidth}
    \includegraphics[width=\textwidth]{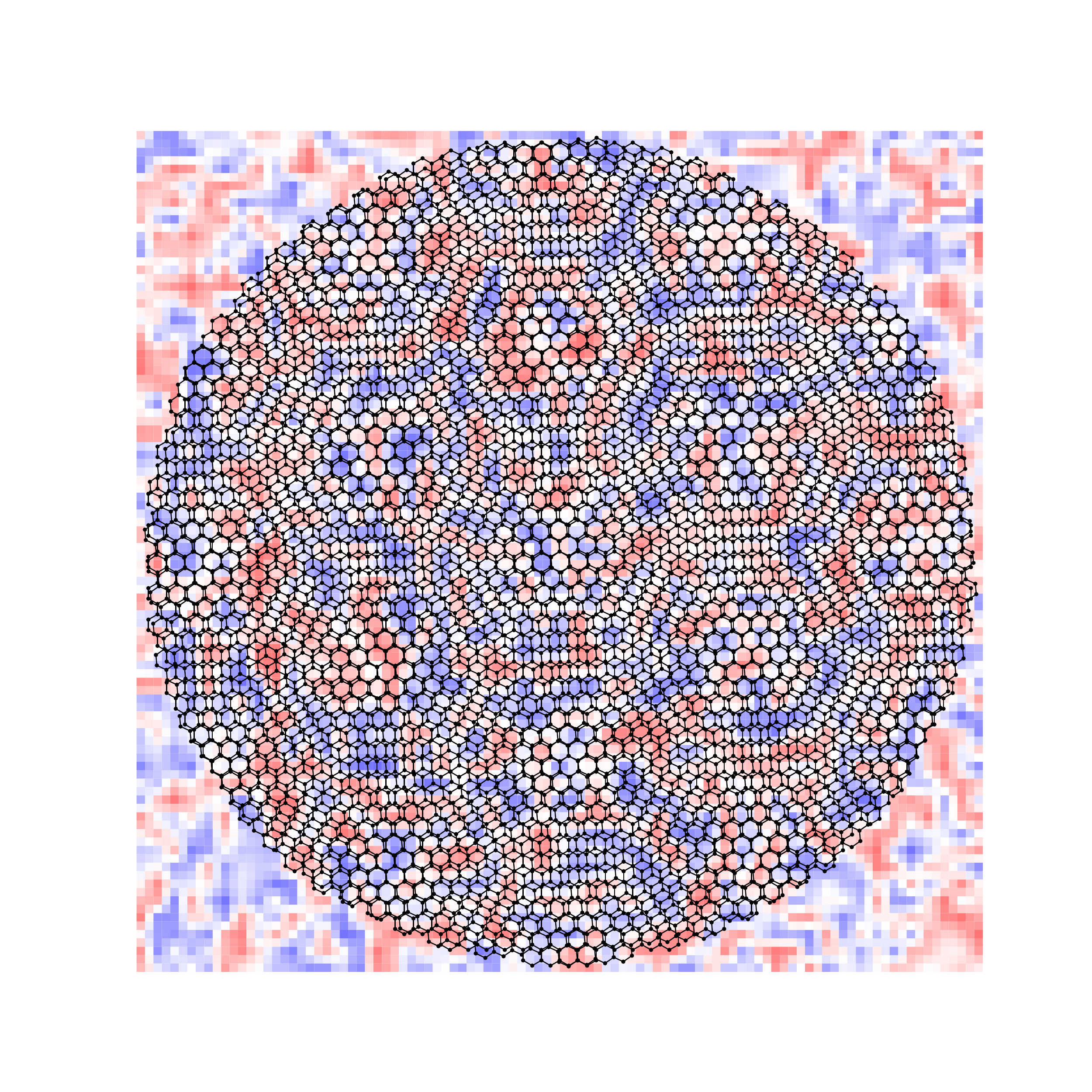}
    \begin{picture}(0,0)
    {\def\unitlength{} \put(-0.5\textwidth,\textwidth){$(a)$}}
    \end{picture}
    \phantomsubcaption{\label{TBG}}
    \end{minipage}
    \begin{minipage}{0.49\linewidth}
    \includegraphics[width=\textwidth]{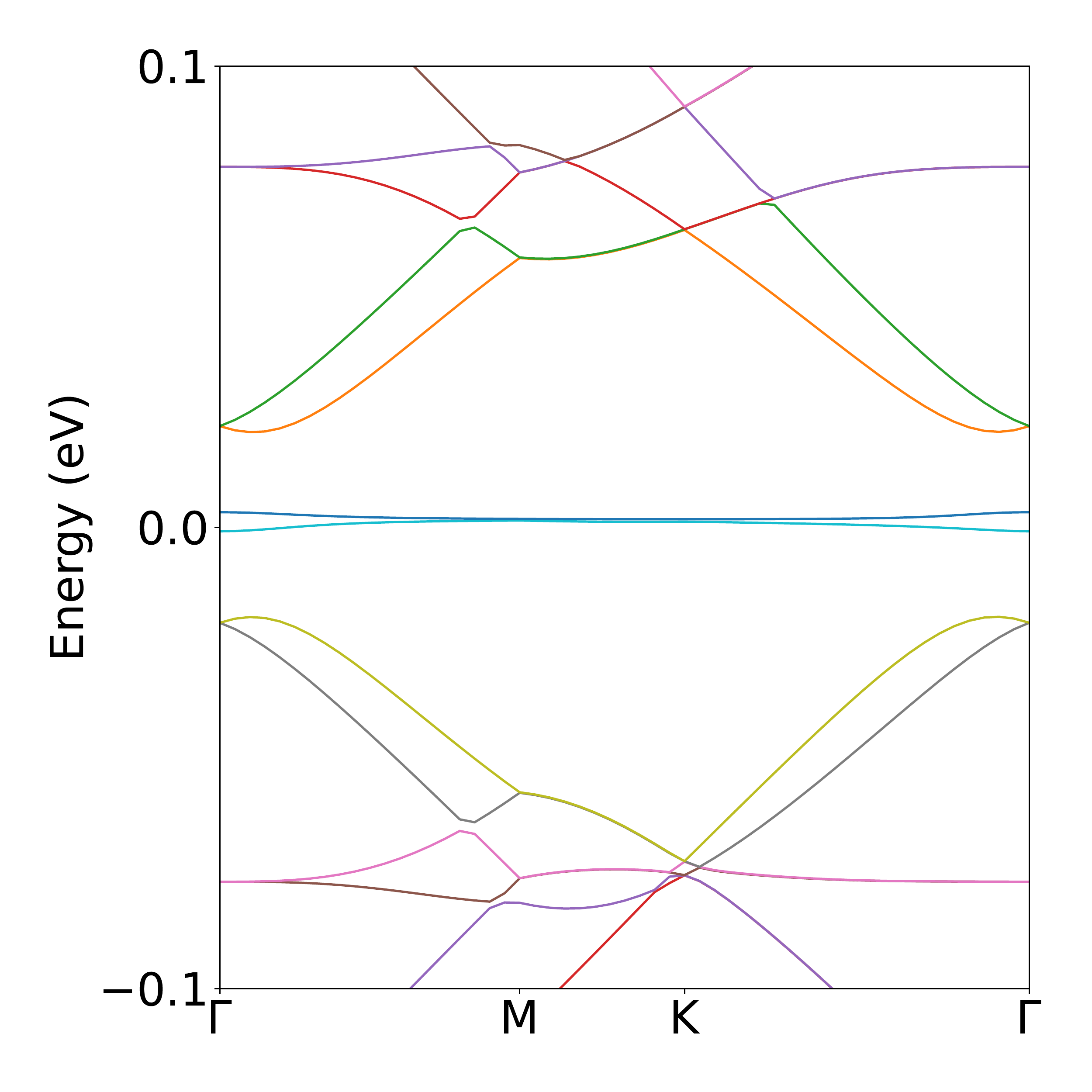}
    \begin{picture}(0,0)
    {\def\unitlength{} \put(-0.5\textwidth,\textwidth){$(b)$}}
    \end{picture}
    \phantomsubcaption{\label{energy}}
    \end{minipage}
    \caption{
Panel \protect\subref{TBG} displays the TBG with the random chemical potential in the background. Electrons fill puddles of random shapes, causing randomness of interactions between flatband states and the emergence of weakly coupled SYK bundles.
 The twisted angle in this figure is chosen to be 5\textdegree{} for illustrative purposes.
Panel \protect\subref{energy} shows the energy spectrum of the non-interacting TBG without the disorder (\cref{tight}) at magic angle $\theta=1.1^{\circ}$ with optimized parameters. The gaped flatbands are located around zero energy.
$\Gamma$ is the center of the Brillouin zone, $M$ is the center of an edge, and $K$ is the corner.
}
\end{figure}
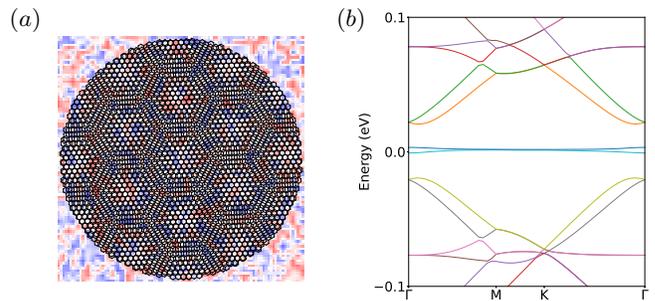

Our considered disorder type originates from the disordered  substrate. The latter induces hills and valleys of random potential with electrons forming paddles of random shapes\cite{skinner2014coulomb}. Although we did the simulation quantitatively with the Dirac system (the three-dimensional Dirac metal examples are Na$_3$Bi\cite{liu2014discovery} or Cd$_3$As$_2$\cite{borisenko2014experimental}), we expect widely used hexagonal Boron Nitrite substrate to exhibit similar disorder-induced effects because of the possibility of having a precise control over carriers and gapless excitations in the system.
Thus, we first discuss the model describing the DTBG system.
The thermodynamics of the DTBG is then compared to the quantum chaotic systems particularly with the maximally chaotic SYK model.
We explore the quantum chaotic dynamics at short and long times and demonstrate quantum scrambling dynamics of the DTBG with a strong disorder, $w$ (see the model description below for the exact definition).
The Gaussian orthogonal ensemble dominates the late-time statistics of the DTBG. We computed the $OTOC$ at short time scales and the extracted low-temperature Lyapunov exponent. We show that it scales linearly with temperature as $\lambda = 2\pi \alpha T$ where $\alpha \simeq 0.56$, indicating that the DTBG is a non-FL lacking quasiparticles. This also agrees with the experimentally observed linear-in-temperature resistivity in TBG \cite{polshyn2019large,cao2020strange}.
Finally, we address the experimentally observed superconducting behavior of magic-angle TBG that is doped around $n=-3$ electrons per Mori\'e unit cell.
We show that upon weakening of disorder strength, the insulating state and the onset of the superconductivity are manifested by the exponentially decaying $OTOC$.
To be more precise, our studies show that the decay rate of the $OTOC$ is insensitive to the strength of the electron-electron interaction in TBG. This fact implies that in a clean limit, the effects of correlations universally manifest themselves in the onset of the formation of the superconducting gap.
Similar exponential decaying behavior is also observed at $n=0$, indicating the insulating and superconducting phases of TBG are closely related.

Experimentally observable characteristics of the SM phase are (i) absence of the logarithmic zero-bias anomaly in the tunneling density of states; (ii) linear-in-temperature specific heat capacity whose slope is essentially different from that of the FL; (iii) thermoelectric transport typical for disordered metals without quasiparticles\cite{davison2017thermoelectric}. We remind the reader that the logarithmic zero bias tunneling anomaly \cite{rudin1997tunneling,mariani2007zero} in the quasi-ballistic regime can be observed in disordered FL due to canceling interference of quasiparticles' Friedel oscillation at low energy. Such anomaly also exists in systems with smooth density variations\cite{sedrakyan2007zero}. 

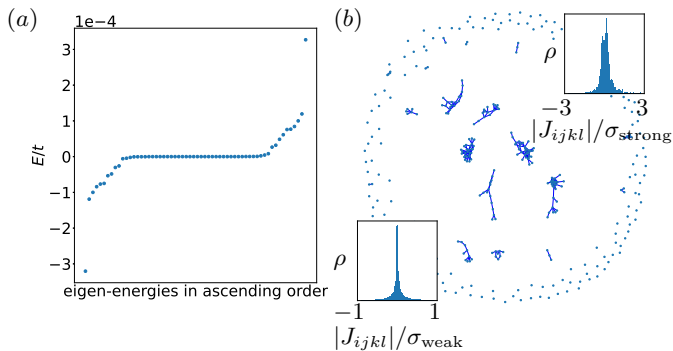
\begin{figure}
    \centering
    \begin{minipage}{0.49\linewidth}
    \includegraphics[width=\textwidth]{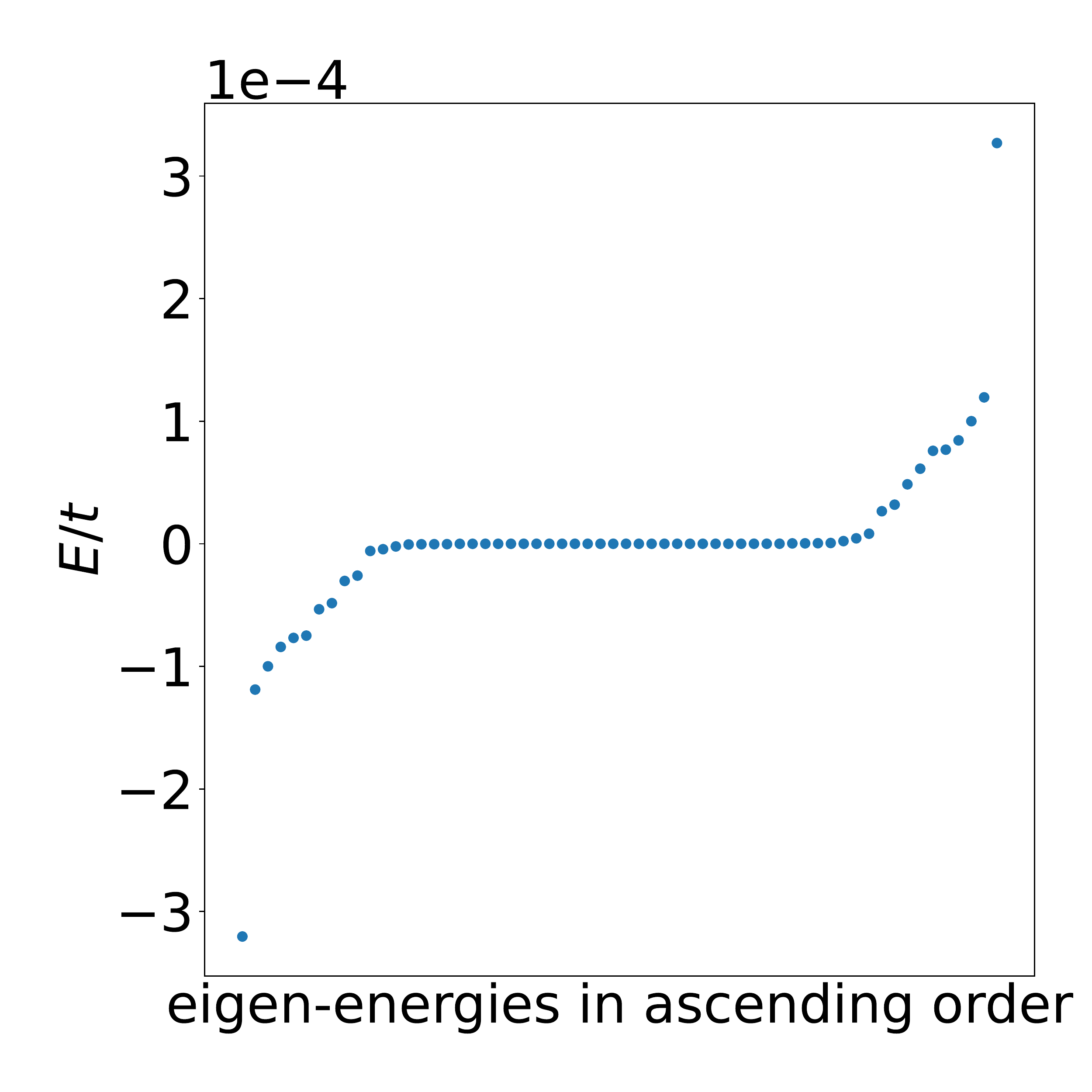}
    \begin{picture}(0,0)
    {\def\unitlength{} \put(-0.5\textwidth,\textwidth){$(a)$}}
    \end{picture}
    \phantomsubcaption{\label{spectrum}}
    \end{minipage}
    \begin{minipage}{0.49\linewidth}
    \includegraphics[width=\textwidth]{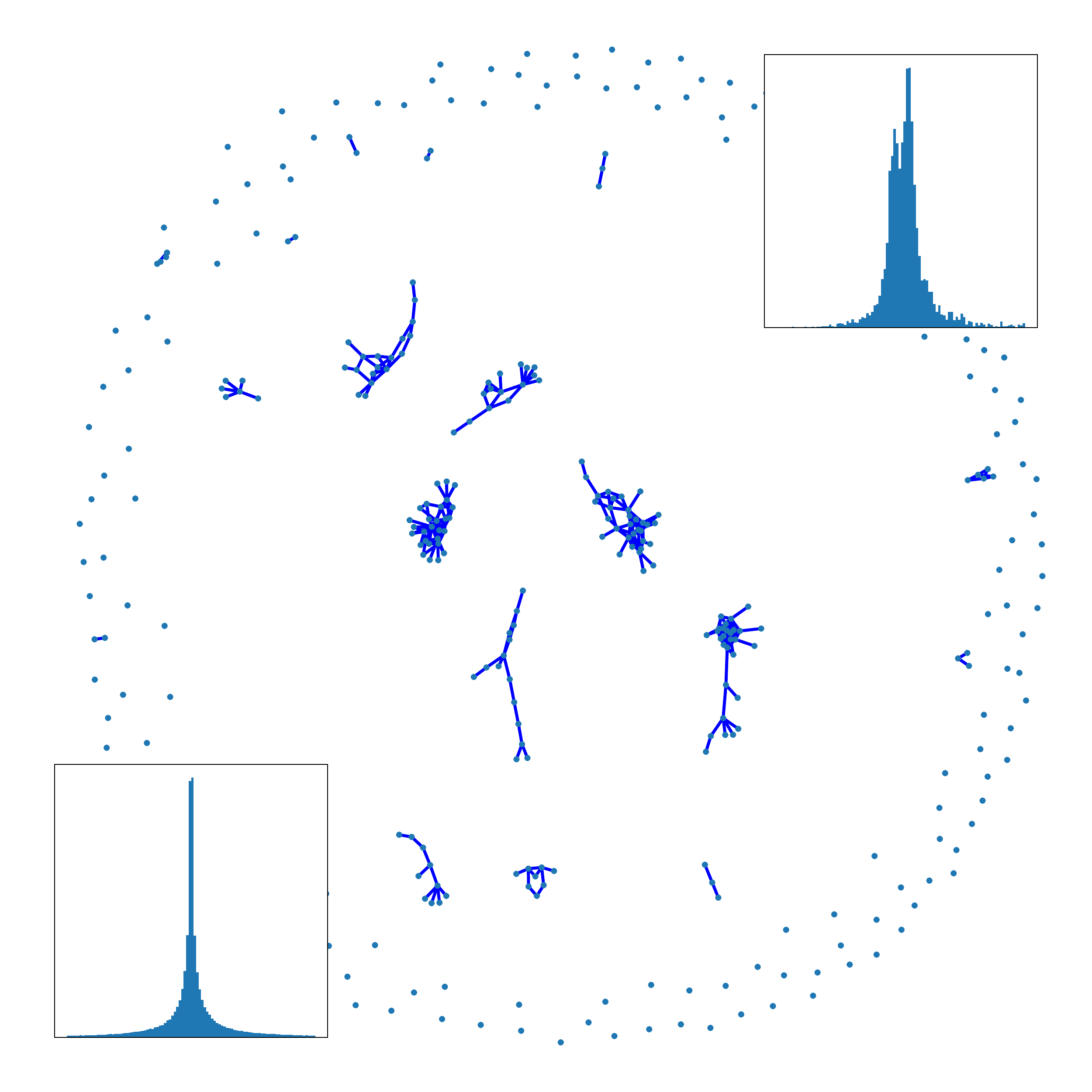}
    \begin{picture}(0,0)
    {\def\unitlength{}
    \put(-0.5\textwidth,\textwidth){$(b)$}
    \put(-0.5\textwidth,0.0\textwidth){$|J_{ijkl}|/\sigma_{\rm weak}$}
    \put(-0.5\textwidth,0.08\textwidth){$-1$}
    \put(-0.2\textwidth,0.08\textwidth){$1$}
    \put(-0.5\textwidth,0.25\textwidth){$\rho$}
    \put(0.12\textwidth,0.65\textwidth){$|J_{ijkl}|/\sigma_{\rm strong}$}
    \put(0.15\textwidth,0.72\textwidth){$-3$}
    \put(0.45\textwidth,0.72\textwidth){$3$}
    \put(0.16\textwidth,0.9\textwidth){$\rho$}
    }
    \end{picture}
    \phantomsubcaption{\label{JTopo}}
    \end{minipage}
    \caption{
Panel \protect\subref{spectrum} shows the energy spectrum around the flatbands of the non-interacting TBG with disorder given by $H_{tt'} + H_{\mu}$ at optimized parameters.
Panel \protect\subref{JTopo} shows the emergence of the weakly connected network of the SYK bundles that is formed by low energy states living in the small energy window around the flatbands. Each pair of low-energy states is associated with a dot. Two dots are connected if the corresponding coupling $|J_{ijkl}| > 0.1 \sqrt{J^2 / 6N^3}$.
The inset on the upper right (lower left) corner shows the distribution of the strong (weak) couplings at the scale of their standard deviation.
The size of each bundle varies from 2 pairs
of low energy states represented by dots to more
than 30 in a single realization. The average bundle size ranges from 11 - 14 dots for different realizations.}
\end{figure}

\section{Model}

The proximity effect of the substrate is demonstrated to be capble of modifying the local potential\cite{pizarro2019internal,liu2021tuning}. Thus, a TBG with coulomb disordered Dirac substrate\cite{skinner2014coulomb} can be described by the Hamiltonian with three terms
\begin{equation}
    H = H_{tt'} + H_{\mu} + H_{I}.
\end{equation}
Here $H_{tt'}$ is the nearest-neighbor hopping between layers and within each layer of the TBG:
\begin{equation} \label{tight}
\begin{aligned}
    H_{tt'} &= -\sum_{\langle \mathbf{r}, \mathbf{r}' \rangle, \sigma, \chi} t a^{\dagger}_{\sigma \chi}(\mathbf{r}) a_{\sigma \chi}(\mathbf{r}') \\
    &\quad - \sum_{\mathbf{r}, \mathbf{r}' , \sigma, \chi} t' e^{-(\mathbf{r} - \mathbf{r}')^2 / r_c^2} a^{\dagger}_{\sigma \chi}(\mathbf{r}) a_{\sigma \bar{\chi}}(\mathbf{r}'),
\end{aligned}
\end{equation}
Operators $a^{\dagger}_{\sigma \chi}(\mathbf{r})$/$ a_{\sigma \chi}(\mathbf{r})$ are the electron creation/annihilation operators with spin $\sigma$ and layer index $\chi$ at the lattice position $\mathbf{r}$. The index $\bar{\chi}$ represents the opposite layer to $\chi$, while $t$ and $t'$ are respectively intralayer and interlayer hopping integrals that acquire positive values, and $r_c$ represents the characteristic length that the local orbital wavefunctions extend. The notation $\langle \mathbf{r}, \mathbf{r}' \rangle$ is adopted to 
represents the nearest-neighbor sites.

In the above expression for the model Hamiltonian,
$H_{I}$ represents the electron-electron interaction
\begin{equation}
\begin{aligned}
    H_{I} = \sum_{\mathbf{r}, \mathbf{r}', \sigma, \chi, \sigma', \chi'} & \frac{1}{2} V_{\sigma, \chi, \sigma', \chi'}(\mathbf{r} - \mathbf{r}') \\
    \cross & a^{\dagger}_{\sigma \chi}(\mathbf{r}) a^{\dagger}_{\sigma' \chi'}(\mathbf{r}') a_{\sigma' \chi'}(\mathbf{r}') a_{\sigma \chi}(\mathbf{r})
\end{aligned}.
\end{equation}
The screened interaction potential is
\begin{equation}
    V_{\sigma, \chi, \sigma', \chi'}(\mathbf{r} - \mathbf{r}') =
    \begin{cases}
    U, \quad \sigma = \bar{\sigma'}, \chi = \chi', \mathbf{r} = \mathbf{r}' \\
    V, \quad \chi = \chi', 0 < |\mathbf{r} - \mathbf{r}'| \leq a \\
    V, \quad \chi \neq \chi', 0 < |\mathbf{r} - \mathbf{r}' + \mathbf{d}| \leq a \\
    0, \quad \mathrm{otherwise}
    \end{cases},
\end{equation}
where $a$ is the distance of the nearest neighbors within a single layer, $\mathbf{d}$ is the perpendicular vector between two layers, $U$ is the on-site interaction for electrons with the different spin, and $V$ is the ``nearest-neighbor'' interaction. The nearest neighbor interactions are restricted to the screening radius $a$ for interlayer and intralayer interactions.

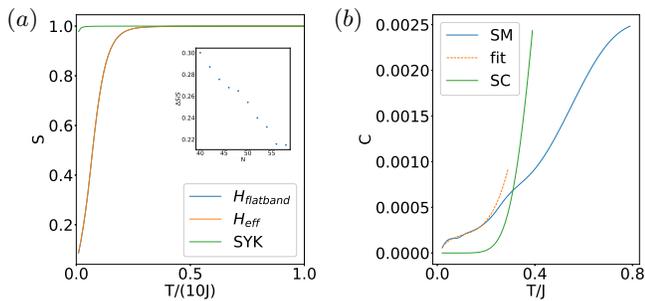
\begin{figure}
    \centering
    \begin{minipage}{0.49\linewidth}
    \includegraphics[width=\textwidth]{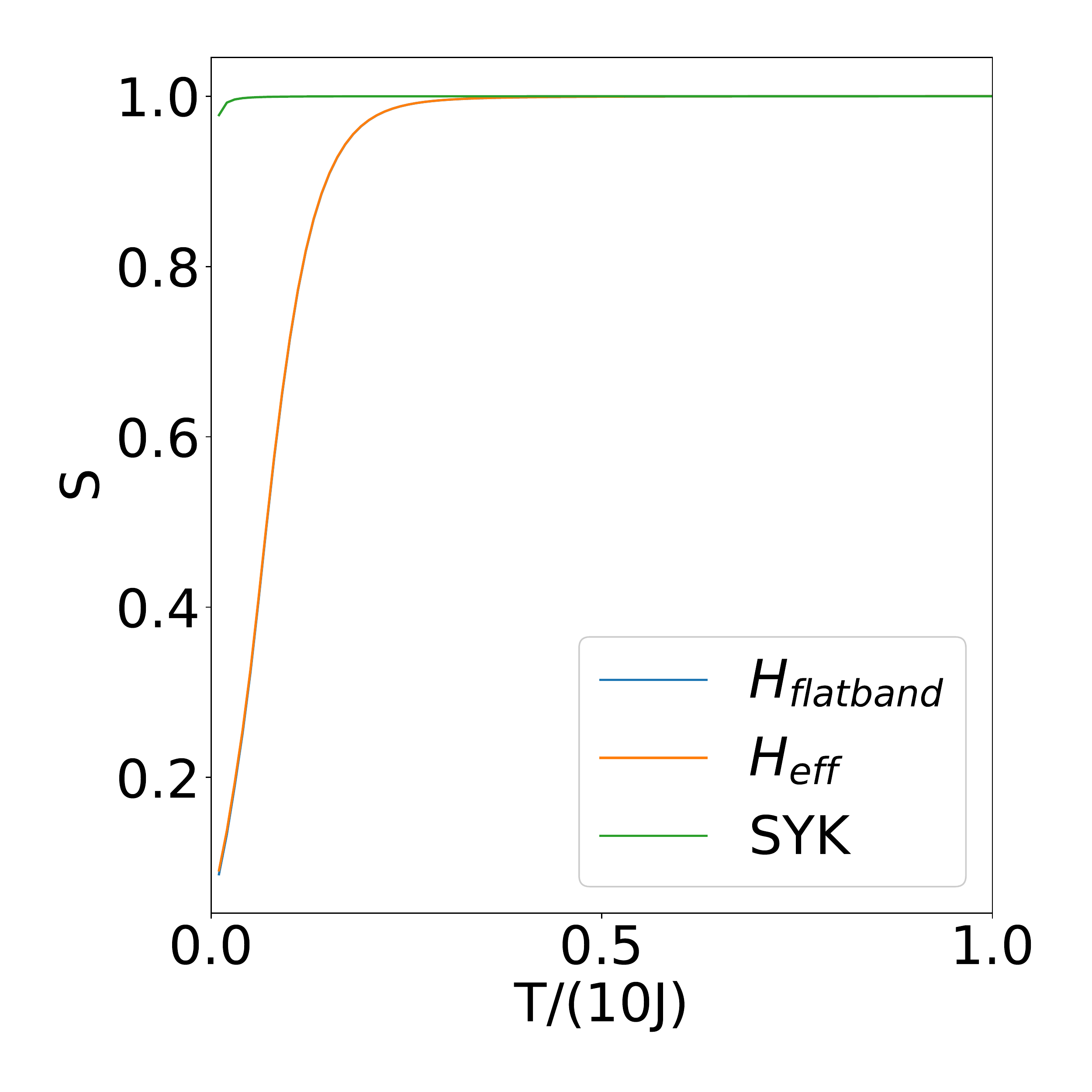}
    \begin{picture}(0,0)
    {\def\unitlength{} \put(-0.5\textwidth,\textwidth){$(a)$}
    \put(0.01\textwidth,0.55\textwidth){\includegraphics[height=0.4\textwidth]{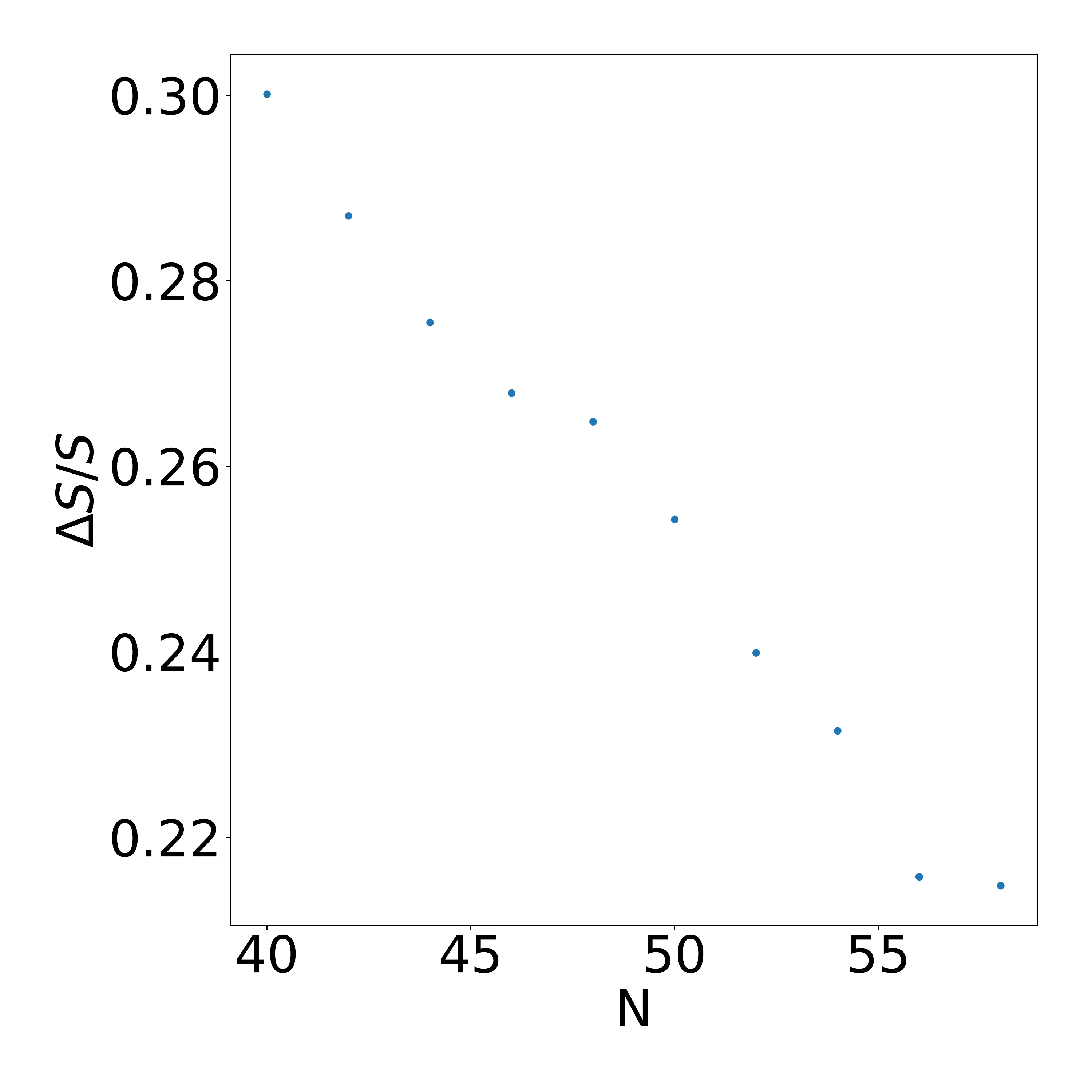}}
    }
    \end{picture}
    \phantomsubcaption{\label{entropy}}
    \end{minipage}
    \begin{minipage}{0.49\linewidth}
    \includegraphics[width=\textwidth]{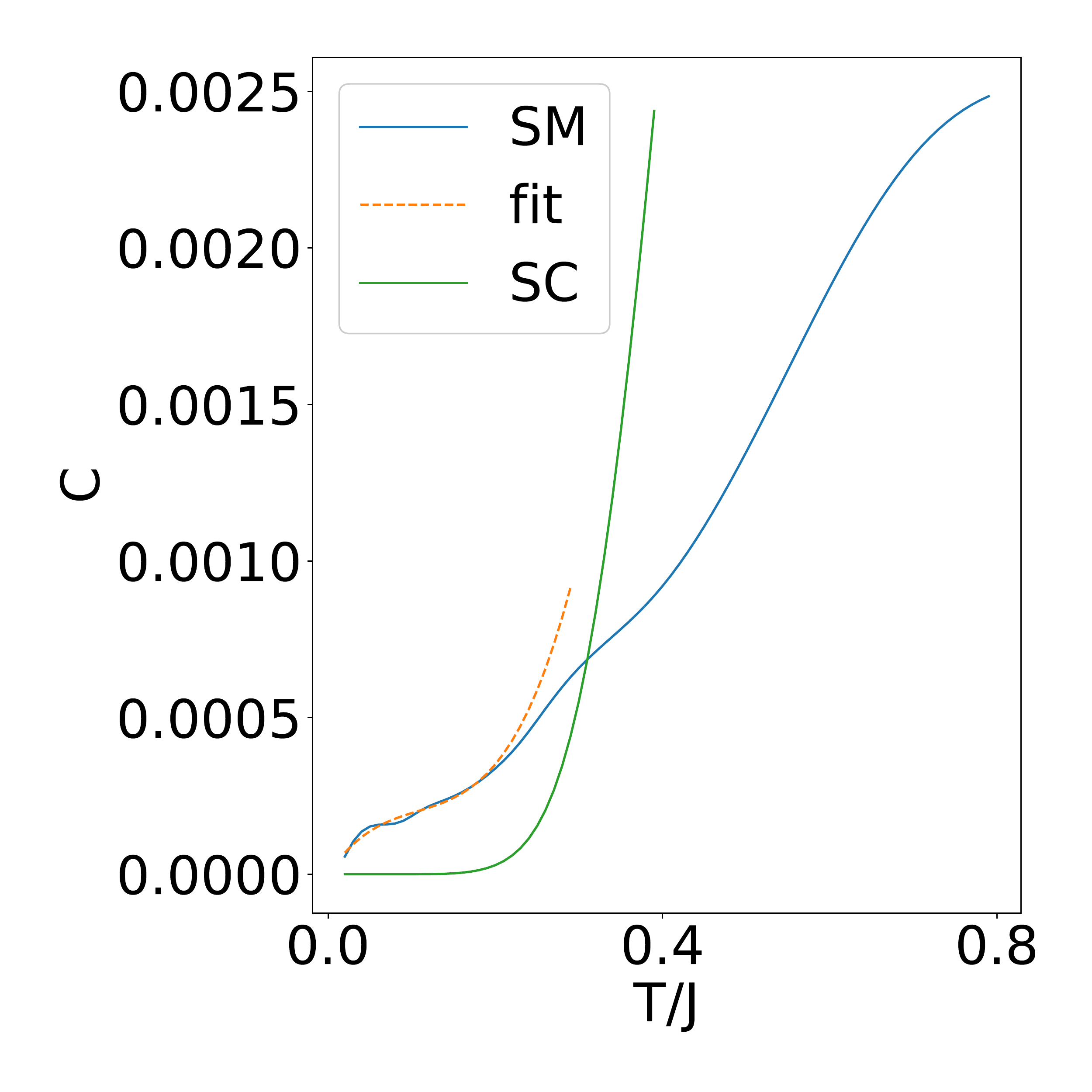}
    \begin{picture}(0,0)
    {\def\unitlength{} \put(-0.5\textwidth,\textwidth){$(b)$}}
    \end{picture}
    \phantomsubcaption{\label{capacity}}
    \end{minipage}
    \caption{
Panel \protect\subref{entropy} shows the entropy of the whole Hamiltonian \cref{eff}, which matches the entropy of the effective Hamiltonian excluding the dispersive part. The result is compared with that of the SYK model.
The inset of panel \protect\subref{entropy} shows the deviation of the entropy with respect to the scaling SYK entropy, $\Delta S = \int_{0}^{10J}dT [S - S_{\rm SYK}(0.1 T)] / (10J)$,
computed as a function of the number of flatband states, $N$. The entropies asymptotically approach each other upon increasing $N$.
Panel \protect\subref{capacity} shows the specific heat of the DTBG. The strange metal (SM) phase is characterized by linear low-temperature behavior. The full curve is fitted by a cubic polynomial function at low temperatures. The exponential decay of $C$ indicates insulating behavior. Upon departing from the charge neutrality, the sharp change in the exponential behavior signals the onset of the superconducting (SC) phase.
        }
\end{figure}

Finally, $H_{\mu}$ represents the 
local random on-site potentials
\begin{equation}
    H_{\mu} = -\sum_{\mathbf{r}, \sigma, \chi} \mu(\mathbf{r}) a^{\dagger}_{\sigma \chi}(\mathbf{r}) a_{\sigma \chi}(\mathbf{r}).
\end{equation}
Here $\mu(\mathbf{r})$ is the mean-field random chemical potential 
approximated by the Gaussian function\cite{conti2021geometry}
\begin{equation}
    \mu(\mathbf{r}) = \sum_i w_{i} e^{-(\mathbf{r} - \mathbf{r}_{i})^2/2\sigma^2},
\end{equation}
where $w_{i}$ is a uniformly distributed random variable on $[-w, w]$ interval, with $w$ representing the energy scale responsible for the strength of the disorder potential.
The parameter $\sigma$ controls the correlation length of the chemical potential, \textit{i.e.}, $\langle \mu(\mathbf{r}) \mu(\mathbf{r}') \rangle \sim e^{-(\mathbf{r} - \mathbf{r}')^2/4\sigma^2}$ in the long wavelength limit. As a result, the position of $\mathbf{r}_i$ does not affect the random potential in the long wavelength limit. This implies that our results do not depend on the structure of the substrate lattice. Here, without loss of generality, we assume that centers $\mathbf{r}_i$ are positioned at the sites of a square lattice in the numerical simulation.

The correlation originated from the Coulomb disordered materials can be estimated $\sigma \approx \left( \frac{1}{2} \right)^{1/6} \frac{1}{8\sqrt{\alpha \pi}} \frac{e^2 n^{1/3}}{\epsilon}$\cite{shklovskii2007simple,skinner2014coulomb}, where $\alpha = \frac{e^2}{4\pi \epsilon \hbar v_{F}}$ is the effective fine-structure constant, $\epsilon$ is the dielectric constant, $n$ is the density of the carriers and $v_F$ is the Fermi velocity of the Dirac system\cite{liu2014discovery,borisenko2014experimental}.
Throughout this paper, we consider the regime when the correlation length of the disordered potential, $\sigma$, is much smaller than the Moir\'e unit cell of the magic-angle twisted TBG. A particular realization of the random chemical potential is shown in \cref{TBG}. The radius of the TBG $R = \frac{\sqrt{3} a}{\theta}$ corresponds to about $19000$ unit cells for each layer. 

\section{Projection onto low energy states: the emergence of the strange metal}

The optimized parameters ensuring the flatness of the bands of $H_{tt'}$ with minimal variance are: $\theta = 1.1 ^{\circ}$, $t' = 0.57 t$, and $r_c = 0.6 \sqrt{3}a$. For these parameters, the spectrum of TBG is shown in \cref{energy}. However, in the presence of the disorder potential with the strong strength chemical potential $H_{\mu}$, that is when the typical energy corresponding to $H_{\mu}$ is of the same order as that of the hopping $H_{tt'}$, the flatbands are disturbed but generally remain localized, as shown in \cref{spectrum}.
In the numerical simulation, the disorder is chosen to be in the range $0.01 t \lesssim w \lesssim t$, which determines the strong disordered phase leading to the chaotic SM phase. In the simulation, we used the states with energy equal to the flatband energy and also the states within the small window of order of $\pm 10^{-4} t$. The number of such low energy states is $58$ in the present simulation. For the weak disordered phase, $w = 0.001t$, we obtained system behavior at $n=0$ (charge neutrality) and $n=-3$ electrons per Mori\'e unit cell, which is consistent with the onset of superconductivity preceding the superconducting phase at higher doping.

We adopt the approach of flatband projection developed and applied in Refs.~\onlinecite{chen2018quantum,wei2021optical} in the context of SYK physics and project the Hamiltonian $H$ onto the low energy subspace where states live inside a small energy window around the flatband, obtaining the low energy effective Hamiltonian. The result is a sum of the interacting Hamiltonian, $H_{\mathrm{flatband}}$, of disperseless states and the dispersive Hamiltonian, $H_{\mathrm{dispersive}}$. The latter is the bilinear part of the effective Hamiltonian $H_{\mathrm{eff}}$ due to the ``non-flatness'' of the low energy states:
\begin{equation} \label{eff}
    H_{\mathrm{eff}} = H_{\mathrm{flatband}} + H_{\mathrm{dispersive}},
\end{equation}
\begin{equation} \label{flatband}
    H_{\mathrm{flatband}} = \sum_{i<j,k<l} J_{ijkl} c_{i}^{\dagger} c_{j}^{\dagger} c_{k} c_{l},
\end{equation}
where $c_{i}^{\dagger}$ ($c_{i}$) is creation (annihilation) operator of the states with energies within a small energy interval around the flatbands. The coupling constants are given by
\begin{equation} \label{Jijkl}
\begin{aligned}
    J_{ijkl} = & \sum_{\mathbf{r}, \mathbf{r}', \sigma, \chi, \sigma', \chi'} V_{\sigma, \chi, \sigma', \chi'}(\mathbf{r} - \mathbf{r}') \\
    &\quad \cross \left[ \phi^*_{i; \sigma, \chi}(\mathbf{r}) \phi^*_{j; \sigma', \chi'}(\mathbf{r}') - \phi^*_{j; \sigma, \chi}(\mathbf{r}) \phi^*_{i; \sigma', \chi'}(\mathbf{r}') \right] \\
    &\quad \cross \left[ \phi_{k; \sigma', \chi'}(\mathbf{r}') \phi_{l; \sigma, \chi}(\mathbf{r}) - \phi_{l; \sigma', \chi'}(\mathbf{r}') \phi_{k; \sigma, \chi}(\mathbf{r}) \right]
\end{aligned},
\end{equation}
where $\phi_{i; \sigma, \chi}(\mathbf{r})$ is the wavefunction of the $i$-th nearly-flatband state.

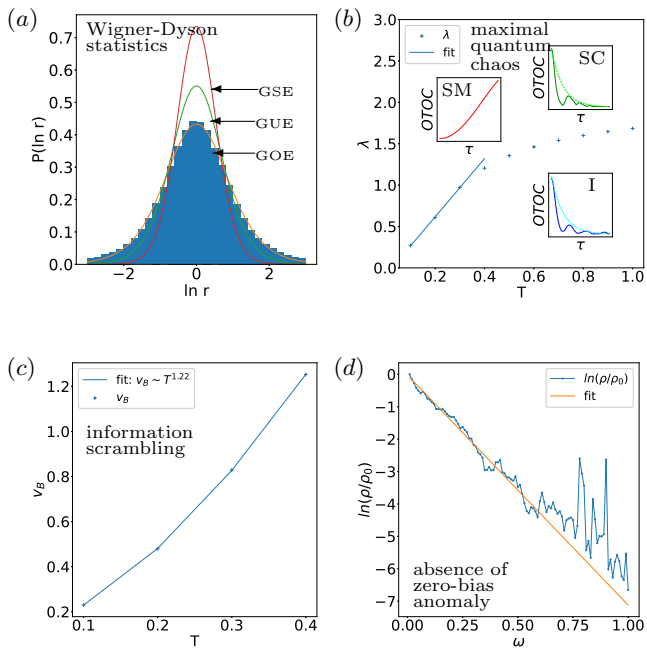
\begin{figure}
    \centering
    \begin{minipage}{0.49\linewidth}
    \includegraphics[width=\textwidth]{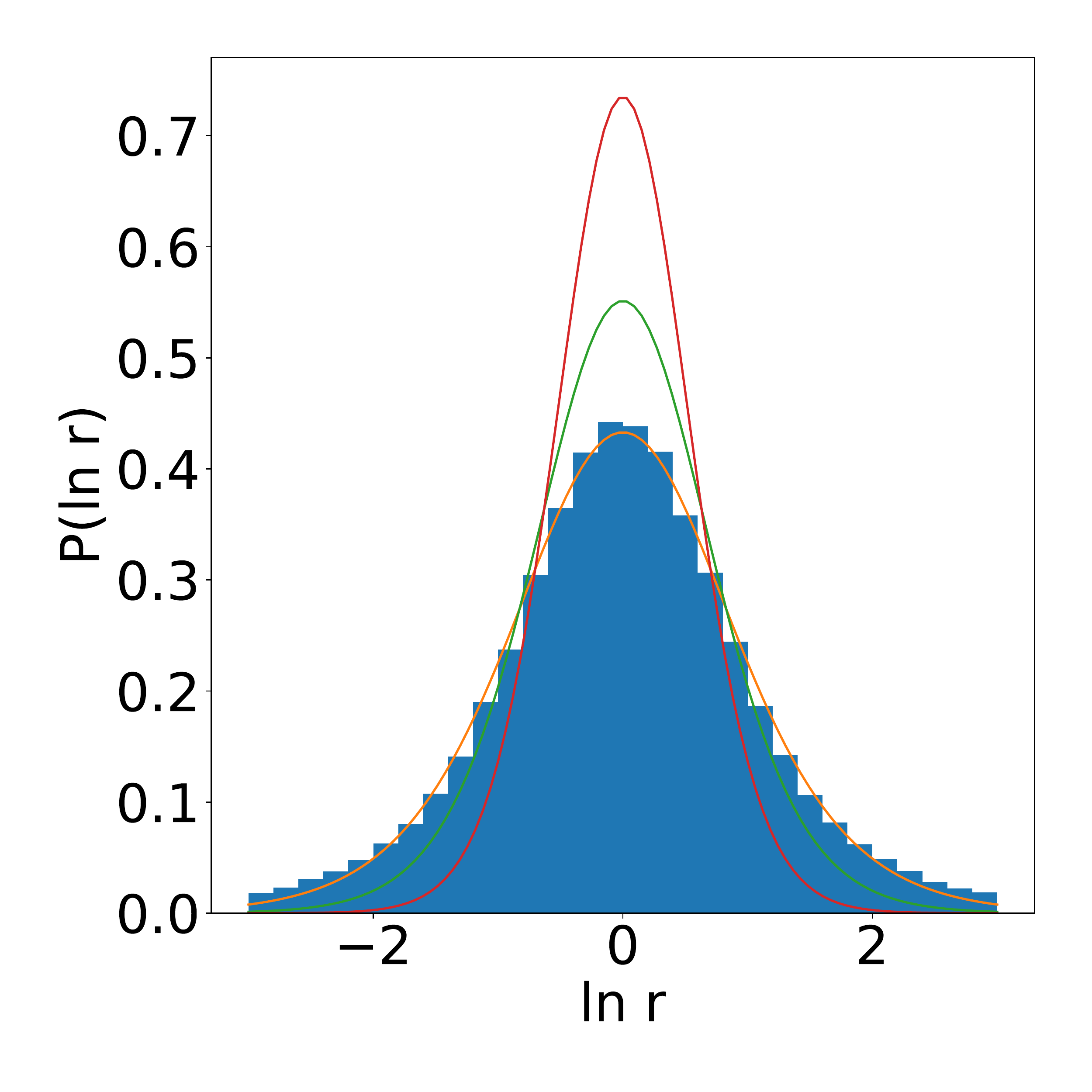}
    \begin{picture}(0,0)
    {\def\unitlength{} \put(-0.5\textwidth,\textwidth){$(a)$}}
    {\def\unitlength{} \put(0.29\textwidth,0.78\textwidth){\text{\tiny GSE}}}
    {\def\unitlength{} \put(0.29\textwidth,0.8\textwidth){\vector(-1,0){18pt}}}
    {\def\unitlength{} \put(0.28\textwidth,0.68\textwidth){\text{\tiny GUE}}}
    {\def\unitlength{} \put(0.28\textwidth,0.7\textwidth){\vector(-1,0){17pt}}}
    {\def\unitlength{} \put(0.28\textwidth,0.58\textwidth){\text{\tiny GOE}}}
    {\def\unitlength{} \put(0.28\textwidth,0.6\textwidth){\vector(-1,0){16pt}}}
    {\def\unitlength{} \put(-0.25\textwidth,0.97\textwidth){\scriptsize Wigner-Dyson}
    {\def\unitlength{} \put(-0.25\textwidth,0.92\textwidth){\scriptsize statistics}}}
    \end{picture}
    \phantomsubcaption{\label{GEO}}
    \end{minipage}
    \begin{minipage}{0.49\linewidth}
    \includegraphics[width=\textwidth]{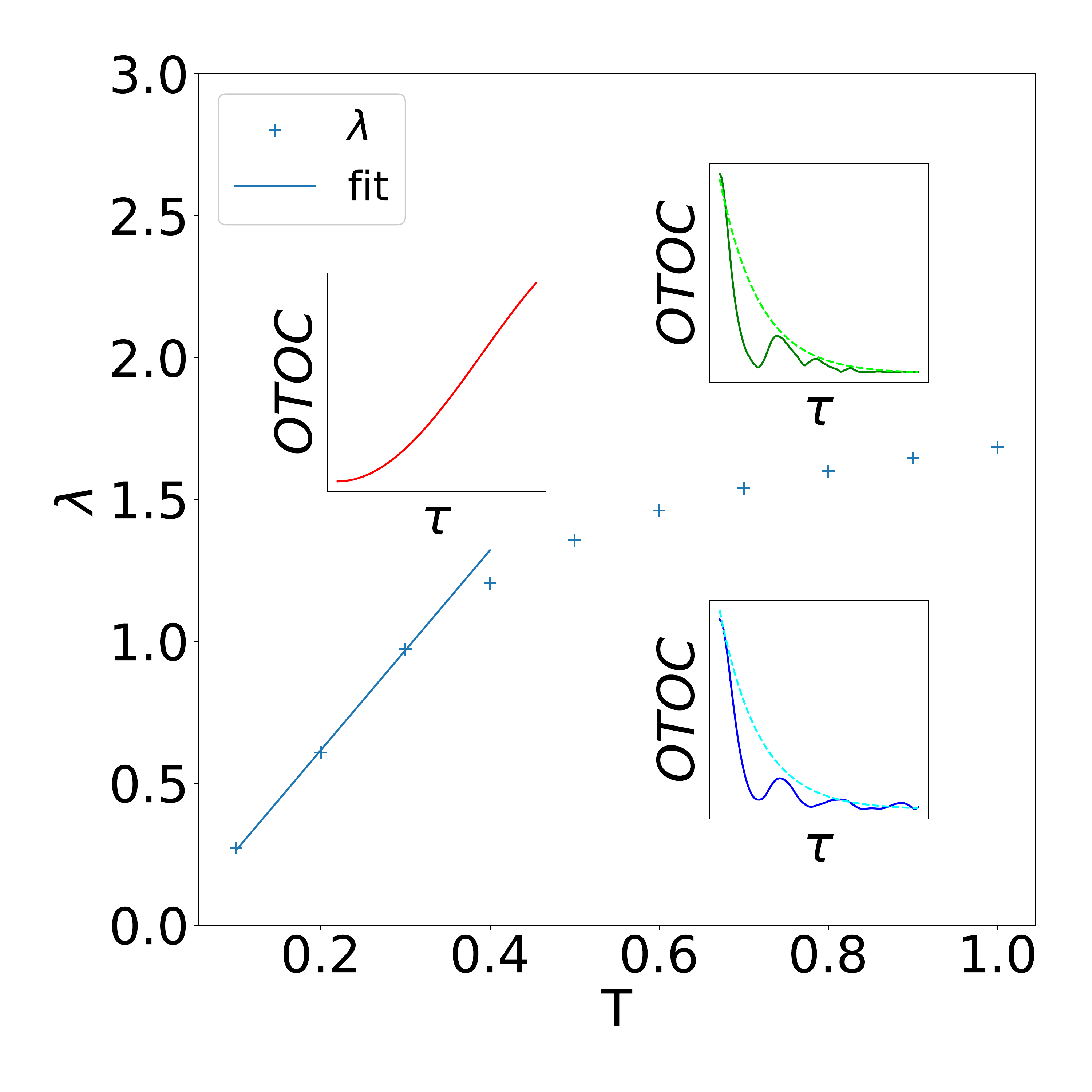}
    \begin{picture}(0,0)
    {\def\unitlength{} \put(-0.5\textwidth,\textwidth){$(b)$}}
    {\def\unitlength{} \put(-0.07\textwidth,0.97\textwidth){\scriptsize maximal}}
    {\def\unitlength{} \put(-0.07\textwidth,0.92\textwidth){\scriptsize quantum}}
    {\def\unitlength{} \put(-0.07\textwidth,0.86\textwidth){\scriptsize chaos}}
    {\def\unitlength{} \put(-0.16\textwidth,0.78\textwidth){\scriptsize SM}}
    {\def\unitlength{} \put(0.3\textwidth,0.48\textwidth){\scriptsize I}}
    {\def\unitlength{} \put(0.27\textwidth,0.88\textwidth){\scriptsize SC}}
    \end{picture}
    \phantomsubcaption{\label{Lyapunov}}
    \end{minipage}
    \begin{minipage}{0.49\linewidth}
    \includegraphics[width=\textwidth]{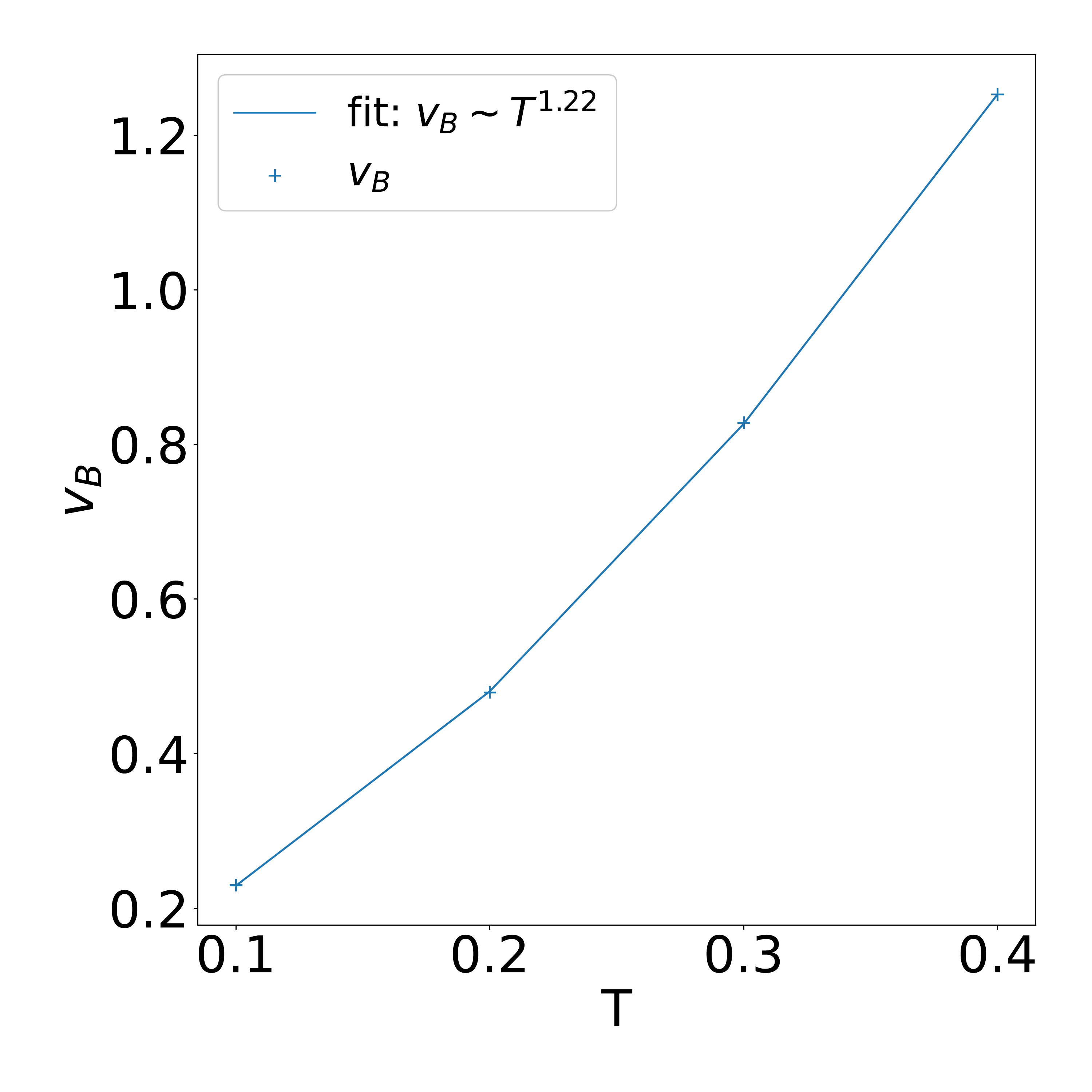}
    \begin{picture}(0,0)
    {\def\unitlength{} \put(-0.5\textwidth,\textwidth){$(c)$}}
    {\def\unitlength{} \put(-0.25\textwidth,0.8\textwidth){\scriptsize information}}
    {\def\unitlength{} \put(-0.25\textwidth,0.75\textwidth){\scriptsize scrambling}}
    \end{picture}
    \phantomsubcaption{\label{vB}}
    \end{minipage}
    \begin{minipage}{0.49\linewidth}
    \includegraphics[width=\textwidth]{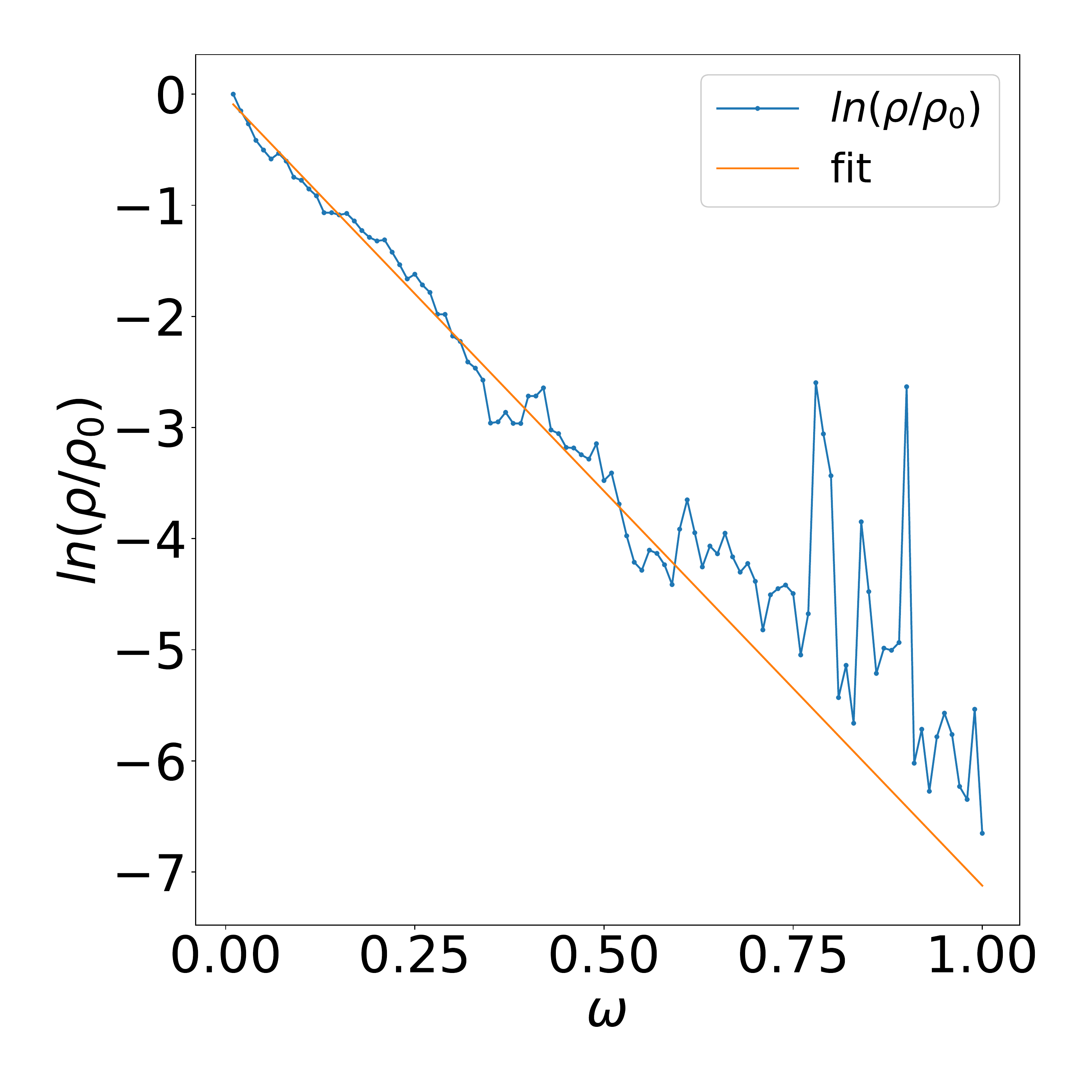}
    \begin{picture}(0,0)
    {\def\unitlength{} \put(-0.5\textwidth,\textwidth){$(d)$}}
    {\def\unitlength{} \put(-0.25\textwidth,0.37\textwidth){\scriptsize absence of}}
    {\def\unitlength{} \put(-0.25\textwidth,0.32\textwidth){\scriptsize zero-bias}}
    {\def\unitlength{} \put(-0.25\textwidth,0.27\textwidth){\scriptsize anomaly}}
    \end{picture}
    \phantomsubcaption{\label{rho}}
    \end{minipage}
    \caption{
Panel \protect\subref{GEO} shows the histogram of $r$. The solid lines with narrower widths correspond to the Gaussian ensemble with orthogonal, unitary, and symplectic matrices.
Panel \protect\subref{Lyapunov} shows the Lyapunov exponent extracted from the $OTOC$ and its low-temperature trend. The left (right) inset shows the $OTOC$ with the strong (weak) disorder supporting the SM (onset of SC) phase.
The SM phase is observed at strong disorder, $w = t$, and the SC and I phases are observed at weak disorder, $w = 0.001t$.
Panel \protect\subref{vB} shows the algebraic increase of $v_B$ at low temperatures.
Panel \protect\subref{rho} depicts the logarithm of the averaged local density of states at low energy, exhibiting the absence of the zero-bias anomaly.
    }
\end{figure}


In the numerical simulation, we choose $V/U = 0.5$, consistent with Ref.~\onlinecite{tancogne2020parameter}. Because most of the low energy states living in the small energy window discussed above are non-extended, the majority of the couplings $J_{ijkl}$ are small. After filtering out these near-zero couplings, the numerical result shows that the remaining couplings are nearly-Gaussian with variance $J^2 / 6N^3$($N$ is the number of flatband states). \Cref{JTopo} depicts how randomly interacting bundles with couplings above a certain threshold emerge. Here each dot represents a pair of low energy states with energies in the small energy window. Two dots are connected if the corresponding coupling, $|J_{ijkl}|$, is larger than $0.1 \sqrt{J^2 / 6N^3}$.
The low energy states are bundled in the dual graph, indicating the sparsely connected feature of the couplings. We notice that such a bundled structure resembles the spinful version of the SYK model in Ref.~\onlinecite{lantagne2021superconducting} with superconducting instabilities and generalizes it to the multi-dots situation.
The size of each bundle varies from 2 pairs of low energy states represented by dots (namely, 4 low energy states with strong random interactions) to more than 30 in a single realization of the numerical simulation of the parameters like $\theta$, $t$, $t'$, $r_c$, etc. For different realizations, the average bundle size ranges from 11 - 14. The finite size effects in flatband realizations of the SYK model are studied in Refs.~\onlinecite{pikulin2017black,wei2021optical}. The results show significant suppression of finite size effects with increasing number $N$. Essentially, for $N > 16$, the behavior of the system is qualitatively and quantitatively similar to the $N \to \infty$ SYK limit.
Although we do not explicitly show in \cref{JTopo} how the SYK bundles are connected through weak couplings($|J_{ijkl}| < 0.1 \sqrt{J^2 / 6N^3}$), these connections do exist, whose distribution is shown in the inset to \cref{JTopo}. The standard deviation of the distribution is $\sigma_{\rm weak} = 0.02 \sigma_{\rm strong}$ confirming that the chosen splitting scale well separates the two types of couplings. Generally, weak connections play an important role for such bundled systems with zero-energy modes to be conductors\cite{day2020nature}. The on-site and nearest-neighbor interaction ratio $V/U$ does not affect the ratio between strong and weak SYK couplings. The reason is that our numerical results already give an average bundle size much larger than the on-site degrees of freedom where the $V/U$ is a control parameter.

Based on the observed energy scale separation of random couplings $J_{ijkl}$, we summarize the non-dispersive part of the Hamiltonian by separating the weak and strong couplings from each other:
\begin{equation}
\begin{aligned}
    H_{\mathrm{flatband}} =& \sum_{\mu} \sum_{i_\mu<j_\mu,k_\mu<l_\mu} J_{i_\mu j_\mu k_\mu l_\mu}^\mu c_{\mu,i_\mu}^{\dagger} c_{\mu,j_\mu}^{\dagger} c_{\mu,k_\mu} c_{\mu,l_\mu} \\
    &+ \sum_{\mu,\nu} \sum_{i_\mu<j_\mu,k_\nu<l_\nu} J_{i_\mu j_\mu k_\nu l_\nu}^{\mu\nu} c_{\mu,i_\mu}^{\dagger} c_{\mu,j_\mu}^{\dagger} c_{\nu,k_\nu} c_{\nu,l_\nu},
\end{aligned}
\end{equation}
where $c_{\mu,i_\mu}^{\dagger}$ ($c_{\mu,i_\mu}$) is the creation (annihilation) operator of the $i$-th state with energy in the small interval around the flatbands in the respective bundle with index $\mu=1 \ldots M$. Here $M$ is the number of bundles in a given realization. The parameters $J_{i_\mu j_\mu k_\mu l_\mu}^\mu$ represent the strong SYK-like couplings that are random and uncorrelated with zero mean value. Parameters $J_{i_\mu j_\mu k_\nu l_\nu}^{\mu\nu}$ are the weak couplings between bundles $\mu$ and $\nu$. We checked numerically that the parameters $J_{i_\mu j_\mu k_\mu l_\mu}^\mu$ and $J_{i_\mu j_\mu k_\nu l_\nu}^{\mu\nu}$ are zero-mean random variables.
, \textit{i.e.},
\begin{equation}
\langle J_{i_\mu j_\mu k_\mu l_\mu}^\mu \rangle = 0, \quad \langle J_{i_\mu j_\mu k_\nu l_\nu}^{\mu\nu} \rangle = 0.
\end{equation}
In the SYK model, one expects random couplings to be independent (uncorrelated). However, slight correlations between random couplings in a random interacting system like this do not affect the central physics of the SYK model. Namely, the fact that the entropy of the system approaches finite value at low temperature is not affected by weak correlations between random couplings\cite{pikulin2017black,wei2021optical}. What is essential is the fact of having an AdS dual description of the theory, which can be achieved in the emergence of reparametrization symmetry. To have fully independent couplings, one expects $\delta$ correlated $J$s:
\begin{equation} \label{Jcorr}
    \begin{aligned}
        \langle J_{i_\mu j_\mu k_\mu l_\mu}^\mu J_{i_\nu j_\nu k_\nu l_\nu}^\nu \rangle &\propto \delta_{\mu\nu} \delta_{i_\mu i_\nu} \delta_{j_\mu j_\nu} \delta_{k_\mu k_\nu} \delta_{l_\mu l_\nu}, \\
        \langle J_{i_\mu j_\mu k_\nu l_\nu}^{\mu\nu} J_{i_\rho j_\rho k_\sigma l_\sigma}^{\rho\sigma} \rangle &\propto \delta_{\mu\nu} \delta_{\rho\sigma} \delta_{i_\mu i_\nu} \delta_{j_\mu j_\nu} \delta_{k_\mu k_\nu} \delta_{l_\mu l_\nu} \\
        &\quad \delta_{i_\rho i_\sigma} \delta_{j_\rho j_\sigma} \delta_{k_\rho k_\sigma} \delta_{l_\rho l_\sigma}.
    \end{aligned}
\end{equation}
This is likely not the case in our simulations. However, we would like to emphasize that this property
in \cref{Jcorr}
is not crucial for the observation of SYK-related strange metal behavior of the system. 


\section{Thermodynamics}

Using stochastic expansion\cite{weisse2006kernel}, we numerically tackle the DTBG whose system size is large enough to have several Moir\'e  cells.
In the stochastic expansion, the trace is replaced by the average of random vectors. Ref.~\onlinecite{weisse2006kernel} has shown that the average converges to the expectation value sandwiched in the true state, and the fluctuation is proportional to $1/\sqrt{NR}$, where $N$ is the dimension of flatband Hamiltonian \cref{eff}, and $R$ is half of the degrees of freedom of the random vectors, allowing access to physical quantities for DTBG of intermediate size.

In the simulation, we choose the number of electrons per Mori\'e unit cell to be $n=0$ and $n=-3$. Here the insulating behavior of the TBG is observed in reasonably clean samples \cite{cao2018unconventional,lu2019superconductors}. This behavior is consistent with our numerical simulation that gapped phase is observed when the disorder $w$ becomes small.



We start with the thermodynamic entropy
\begin{equation}
S(\beta) = -\tr \rho \ln \rho, ~\mathrm{where}~  \rho = \frac{e^{-\beta H_{\mathrm{eff}}}}{Z}.
\end{equation}
$H_{eff}$ is given by \cref{eff} and $Z$ is the partition function.
The result of the simulation is
shown in \cref{entropy}.
Choosing the $10J$ (we remind the reader that $J^2/6N^3$ is the variance of the effective coupling \cref{Jijkl}) as the scale for the temperature, we observe that
the entropy of the effective Hamiltonian \cref{eff} of DTBG
decreases to a small value at temperatures much higher compared with the SYK model\cite{fu2016numerical} with the same value of the
parameter $J$. Despite the energy scale, the entropies show similarity in the shape\cite{wei2021optical}, which leads to a conjecture $S(T) = S_{\rm SYK}(\zeta T)$, where $\zeta \approx 0.1$.
This indicates that, unlike the maximally chaotic SYK model, the ground state of the DTBG is not as largely degenerate, resulting from the bundling structure of the
couplings $J_{ijkl}$. We also checked that the dispersive part in the Hamiltonian \cref{eff} does not affect the described picture, hence can be safely neglected. We do so in the remaining part of the paper based on the observation that the energy scale corresponding to the dispersive part of the TBG is small enough compared with the interaction. A similar effect was observed in Ref.~ \onlinecite{pikulin2017black}, where it was shown that because of the similar separation of the energy scales, the two body terms could be neglected.

At the superconducting phase, the specific heat decays exponentially as $e^{-T/\Delta}$ where $\Delta$ is the superconducting gap. Yet the specific heat of the chaotic phase shows more complicated behavior where the low-temperature part can be fitted by the cubic polynomial $c_1 (T/J) + c_2 (T/J)^2 + c_3 (T/J)^3$. The linearity coefficient,
\begin{equation}
c_1 \simeq 0.04,
\end{equation}
is about one-tenth of the analytical and numerical results for the SYK model\cite{maldacena2016comments,garcia2016spectral}. This also agrees with the scale conjecture about the entropies, $C(T) = \frac{T}{N} \frac{\partial S(T)}{\partial T} = \frac{\zeta T}{N} \frac{\partial S_{\rm SYK}(\zeta T)}{\partial (\zeta T)} = C_{\rm SYK}(\zeta T)$. Remarkably, the specific heat fits better with the polynomial function than a logarithmic form $T \ln(1/T)$, suggesting the specific heat of the SM phase is closer to the SYK model compared with the ``normal'' phase of the superconductor\cite{patel2022universal,michon2022planckian}.

\section{Quantum chaos and non-Fermi liquid behavior}

As observed thermodynamically, the DTBG ground state are not largely degenerate at energy scale $J$, which is a result of the sparsely connected topology of the couplings $J_{ijkl}$. However, the dynamics is determined by the local interacting structure of the bundles.
To understand the long-time dynamical behavior of the DTBG, we compute the level statistics, namely the distribution of the level-spacing parameter $r_n = \frac{E_{n+1} - E_{n}}{E_{n} - E_{n-1}}$ corresponding to energy levels $E_n$.
Here the energy of the $n$-th level can be determined by the min-max theorem\cite{teschl2014mathematical}
$
   E_{n} = \min_{U} \max_{x} \{ \bra{x} H_{\mathrm{eff}} \ket{x} : x \in U, |x| = 1, \dim(U)=n \}
$,
which is a variational method minimizing the $n$-th eigenenergy with in subspace $U$ whose dimension is $n$.

As seen from \cref{GEO}, the distribution of $r_n$ follows the Gaussian orthogonal ensemble. This is because the time-reversal symmetry is respected in our model\cite{you2017sachdev}.
To further understand the short-time behavior of the model, we compute the $OTOC$ of the flatband states,
\begin{equation}
\begin{aligned}
    OTOC(\tau, \beta) = \tr e^{-\frac{\beta H_{\mathrm{flatband}}}{2}} c_{i}^{\dagger}(\tau) c_{i} e^{-\frac{\beta H_{\mathrm{flatband}}}{2}} c_{i}^{\dagger}(\tau) c_{i},
\end{aligned}
\end{equation}
where $c_i^{\dagger}$ ($c_i$) are creation (annihilation) operators of the nearly-flatband states and
$
    c_{i}^{\dagger}(\tau) = e^{i H_{\mathrm{flatband}} \tau} c_{i}^{\dagger} e^{-i H_{\mathrm{flatband}} \tau}
$
is the Heisenberg picture operator.
The Boltzmann factor $e^{-\beta H_{\mathrm{eff}}}$ is distributed in the definition of the $OTOC$ for regularization\cite{lantagne2020diagnosing}. The Lyapunov exponent(\cref{Lyapunov}) is extracted at different temperatures from the exponential growing portion of the $OTOC$ around the scrambling time\cite{maldacena2016bound}. Respectively, scaling of $\lambda$ versus $T$ is linear, and the slope approaches $3.52$ and $3.47$ as $T \to 0$, which is the $\alpha\simeq 0.56$ and $0.55$ portion of the conjectured maximal bound, observed for $n=0$ and $n=-3$.
Similar effects were reported in a non-FL spin-fermion model\cite{tikhanovskaya2022maximal} and Artin systems\cite{babujian2020correlation}. This indicates substantial quantum scrambling present in the DTBG.

To study the butterfly velocity, we turn to the local electronic operators to have the explicit dependence in $\mathbf{r}$. The $OTOC$ now depends on not only time and temperature, but also real-space coordinates of electrons:
\begin{equation}
\begin{aligned}
    OTOC(\tau, \mathbf{r}, \beta) = \tr &e^{-\frac{\beta H_{\mathrm{flatband}}}{2}} a_{\sigma \chi}^{\dagger}(\tau, \mathbf{r}) a_{\sigma \chi}(\mathbf{0}) \\
    \cross &e^{-\frac{\beta H_{\mathrm{flatband}}}{2}} a_{\sigma \chi}^{\dagger}(\tau, \mathbf{r}) a_{\sigma \chi}(\mathbf{0})
\end{aligned},
\end{equation}
where $a_{\sigma \chi}^{\dagger}(\mathbf{r}) / a_{\sigma \chi}(\mathbf{r})$ are the electron creation/annihilation operators with spin $\sigma$ and layer index $\xi$ at the lattice position $\mathbf{r}$, and
$
    a_{\sigma \chi}^{\dagger}(\tau, \mathbf{r}) = e^{i H_{\mathrm{flatband}} \tau} a_{\sigma \chi}^{\dagger}(\mathbf{r}) e^{-i H_{\mathrm{flatband}} \tau}
$
is the corresponding Heisenberg picture operator.

For a given $\tau$, the butterfly velocity $v_B$ can be extracted from the exponential behavior of $OTOC$. \Cref{vB} depicts the actual data and fitted $v_B $ versus temperature showing 
\begin{equation}
v_B \sim T^{\delta}
\end{equation}
with
\begin{equation}
    \delta\simeq
    \begin{cases}
        1.22, \mathrm{for ~} n=0 \\
        1.24, \mathrm{for ~} n=-3
    \end{cases}.
\end{equation}
For FL with dispersion relation $\omega \sim k^z$ at finite temperature $T$, the butterfly velocity defines the velocity of the operator/information spreading, so $v_B = \frac{d\omega}{dk} \sim \omega^{1-1/z} \sim T^{1-1/z}$. However, this argument does not apply to flatband non-FL, even perturbatively. So $\delta > 1$ becomes a signature for breaking FL theory and the absence of quasiparticles.
Because of the absence of quasiparticles, the zero-bias anomaly is missing. The averaged local density of states, $\rho$, is exponentially decaying at small energies as shown in \cref{rho}.

As we can see, the exponents describing the level of quantum chaos, namely the Lyapunov exponent and the exponent in the butterfly velocity, remain almost the same throughout the SM phase. This suggests that the origin of the SM phase is the same in the whole 3-dimensional phase diagram at high and low temperatures/disorder levels.

At weak disorder, $w = 0.001 t$, we observed a negative $\lambda$.
If we keep the ratio between the ``nearest-neighbor'' interaction and the on-site interaction for electrons with the different spin, $V/U$, unchanged while simultaneously increasing $V$ and $U$, the non-dispersive Hamiltonian $H_{\rm flatband}$ will only gain the same factor. Therefore, the decay rate of the $OTOC$ remains the same. This behavior is different from what one has in a correlated insulator, the energy gap of which strongly depends on the interaction strength. Therefore, we conclude that the exponentially decaying $OTOC$, which signals a transition to a gapped state in the clean limit, represents the superconducting state itself or the onset of a superconducting state. Such a result is also consistent with the superconducting Larkin-Ovchinnikov behavior of $OTOC$ and the thermodynamic result discussed in the previous section.
Another possible mechanism of the gap formation is the dynamical symmetry breaking, where the gap can arise because of different symmetry breaking patterns depending on the strength and range of the Coulomb interaction\cite{grinenko2021state}.

\section{Conclusion and open problems}
We investigated the characteristics of the TBG on a disordered substrate or sandwiched in the Coulomb disordered materials. We used stochastic expansion combined with numerical diagonalization. The thermodynamic entropy of DTBG and the SYK model are compared to determine the energy scale and the bundling structure of the coupling $J_{ijkl}$ in the DTBG. The quantum chaos of the DTBG is demonstrated to originate from the emergence of the weakly coupled network of the SYK bundles and follows the Gaussian orthogonal ensemble statistics. The exponential growth of the $OTOC$ probes quantum scrambling in the DTBG. The temperature scaling of the butterfly velocity, $v_B$,  of the system implies anomalous information spreading with $v_B \sim T^{1.22}$ for $n=0$ and $v_B \sim T^{1.24}$ for $n=-3$.
When $T \to 0$, the corresponding Lyapunov exponent has a linear dependence on the temperature, and the slope is $\sim 0.56$ and $\sim 0.55$ of the conjectured upper bound for $n=0$ and $n=-3$ electrons per Mori\'e unit cell respectively. Upon decreasing the strength of the disorder, the OTOC undergoes a transition to the exponentially decaying Larkin-Ovchinnikov behavior inherent to the superconducting phase or onset of the superconducting phase. Simultaneously, the specific heat shows exponential decay, a typical characteristic of gapped systems. These observations suggest that the superconductivity in the magic angle TBG stems from increased couplings between spatially extended SYK bundles of the SM phase. This can be observed if a sharp change of the exponentially decaying $OTOC$s is detected. The superconductivity in such a non-Fermi-liquid scenario can be stabilized upon the emergence of the attractive Hubbard interaction\cite{wang2020sachdev} between flatband states. It would be interesting to see if the interplay between weak Coulomb disorder and strong random interactions may lead to the effective attraction that leads to pairing. Although we focus on the DTBG here, we believe such chaotic non-Fermi liquid behavior exists in other disordered Mori\'e flatband systems, which may stabilize superconducting phases.


Although the numerical results already indicate the non-FL and quantum chaotic behavior of DTBG, the system size is limited. Hence, the finite size effect is not revealed. It would be illuminating to have a low-energy model for the interacting DTBG. To have such a low-energy model, several behaviors should be understood. The first one is how the localization landscape of disordered flatband states that are created by $c_{i}^{\dagger}$ leads to bundling. It would also be critical to pinpoint the microscopic pairing mechanism of the SC phase\cite{volovik2018graphite} and its relation to the flat band\cite{volovik2019flat,khodel1990superfluidity}.

It is also interesting to understand the interplay between disorder and interaction in systems with reduced ``dispersiveness'', which include not only the flatband systems such as DTBG but also the gapless systems with degeneracy on a closed manifold in the momentum space. Examples include the moat band systems studied in Refs.~\onlinecite{sedrakyan2012composite,sedrakyan2014absence,sedrakyan2015spontaneous,sedrakyan2015statistical,wang2022emergent,wei2022chiral,wang2023excitonic}.

The disorder in twist angle, a common source of disorder in TBG, results in variations in the size of the Moir\'e cell. Consequently, it leads to variations in electron density. A straightforward way to model this effect is by considering variations in the onsite potential. The key difference between this model of angular disorder and the residual one discussed in the current work lies in the disorder correlation length. In the former model, the correlation length would be comparable to the Moir\'e length. Our method allows one to account for angular disorder, the interplay between substrate-induced disorder, and interactions with the resulting SM and SYK physics. This is an open and interesting problem for future investigations.




\begin{acknowledgments}
We thank Yi Huang, Alex Kamenev, and Ara Sedrakyan for their valuable discussions. This research was in part supported by the National Science Foundation under Grant No. NSF PHY-1748958.
\end{acknowledgments}

%
\end{document}